# The Lack of Long Range Correlations is a Necessary Condition for a Functional Biologically Active Protein.

E.Sh.Mamasakhlisov, V.F.Morozov & M.S.Shahinian

Yerevan State University, Department of Physics, A.Manoogian st.1, 375049 Yerevan, Armenia

**Abstract.** We study random heteropolymer chain with gaussian distribution of kinds of monomers. The long-range correlations between kinds of monomers were introduce. The mean-field analysis of such heteropolymer indicates the existence of infinite energetic barrier between heteropolymer random coil and frozen states. Thus, the frozen state is kinetically unavailable for the random heteropolymer with power-law correlations in monomers' sequence. The relationship between our results and some early obtained results for the DNA introns sequences are discusse.

**The Lack of Long  Range Correlations is a Necessary Condition**

**for a Functional Biologically Active Protein.**

**1. Introduction.**

The relationship between the sequence and conformation of a protein macromolecule is one of the great unsolved problems in biophysics. At the present time it is widely believed  that functional proteins usually form a single compact three-dimensional structure that corresponds to the global energetic minimum in the conformational space. In recent years the first step to address this question is to study random heteropolymers and compare them with proteins. The fact that even chains with random sequences can  have a unique ground state characterized by frozen path  of polymer chain backbone was first examined in terms of random energy model (REM) [1,2]. The set of  subsequent investigations was carried out on the basis of  "microscopic"  Hamiltonians in which the interactions between pairs of monomers were assumed to be random , independently taken from a Gaussian distribution [3] , or with polymer sequences explicitly  present [4,5,6]. All these models were shown to exhibit freezing phase transition for random chain.

Recently, Shakhnovich and Gutin[7] found that to have such a minimum it is sufficient that an amino acid sequence forms an uncorrelated random sequence. Accompanying these results rise the following question: Is the lack of  long range correlations in  protein sequences a *necessary* condition for a three-dimensional biologically functional structure formation? Here we give some positive reasons for this question.

From this point of view is very interesting that at the recent years some results about long-range (scale-invariant) correlations in non-coding  DNA sequences were obtain  [8]. For example was find that only non-coding DNA sequences exhibits long-range correlations.  Some reports [9,10] support this finding, but other authors [11,12] disagree. For example Voss[10]  recently proposed that coding as well non-coding  DNA sequences display long-range power-law correlations in their base pair sequences.

In the present paper the above mentioned problem will examine for heteropolymer with the quenched random sequence described by the set of random  variables $\sigma_k$ characteristic of the each monomer . In the past [4-6]  the monomer species were considered as independent random variables or as in [13] were examined short range ( with exponential decay ) correlations between kinds of monomers . We are investigate here the folding problem for the random heteropolymer with monomers quenched sequence in presence of long range ( power - law ) correlations .

**2. Model and Mean-Field Theory.**

Let us discuss the heteropolymer chain with a frozen sequence of monomers describing with Hamiltonian that is a function of the monomer coordinates $\{\vec{r}_i\}$. Our model Hamiltonian can be written

$$H = \frac{T}{a^2}\sum_i (\vec{r}_i - \vec{r}_{i+1})^2 + \sum_{i<j} B_{ij}\delta(\vec{r}_i - \vec{r}_j) + \\ + C\sum_{i<j<k} \delta(\vec{r}_i - \vec{r}_j)\delta(\vec{r}_j - \vec{r}_k)$$

(2.1)

where $B_{ij} = B_0 + B\sigma_i\sigma_j$ and $C$ are virial coefficients describing two and three- particle interactions , $\sigma_k$ is variable of specie of k- th monomer, $a$ is the statistical segment length, $T$ is the temperature. We work in units where $k_B = 1$.

Earlier [1-6] this problem in case of $\sigma_k$ statistically independent variables was discuss. In the present paper we are using $\sigma_k$ as random variables with long range correlations , characterized by Gaussian distribution function in form :

$$P\{\sigma\} \propto \exp[-1/2(\vec{\sigma}, \hat{K}^{-1}\vec{\sigma})] \\ \vec{\sigma} = (\sigma_1,\ldots,\sigma_N)$$

(2.2)

where $\hat{K} = \|K_{ij}\|$ is matrix describing correlations in chain sequence :

$$K_{ij} = \langle \sigma_i \sigma_j \rangle_P$$

(2.3)

There are some reasons about existence of the scale-invariant long range correlations in the DNA sequences only in non-coding regions of DNA (introns) (see, e.g. [14] ). The correlation function of the monomers' species in this case has form according to [14]:

$$K(l) \propto l^{\beta-1}$$

(2.4)

where $0 < \beta < 1$ and $l = |i - j|$.

Now we are going to find the free energy $F$. The standard way to derive the partition function of a system with quenched disorder is to employ the replica trick:

$$F = \langle F(\sigma) \rangle_P = -T \lim_{n \to 0} \frac{\partial}{\partial n} \langle Z(\sigma)^n \rangle_P \qquad (2.5)$$

where $\langle \ \rangle_P$ means average over all possible realizations of $\vec{\sigma}$. In these terms averaged value of partition function will come to:

$$\langle Z^n \rangle_P = \int Dr_i^\alpha g(\vec{r}_{i+1}^\alpha - \vec{r}_i^\alpha) \exp(-C \sum_\alpha \sum_{ijk} \delta(\vec{r}_i^\alpha - \vec{r}_j^\alpha)\delta(\vec{r}_i^\alpha - \vec{r}_k^\alpha)) \times$$
$$\exp\left(-B_0 \sum_\alpha \sum_{ij} \delta(\vec{r}_i^\alpha - \vec{r}_j^\alpha)\right) \left\langle \exp(-B \sum_{ij} \sigma_i \sigma_j \sum_\alpha \delta(\vec{r}_i^\alpha - \vec{r}_j^\alpha)) \right\rangle_P. \qquad (2.6)$$

where $\alpha$ is replica indices, $\vec{r}_i^a$ - describes the position of i-th monomer of replica $\alpha$ in three dimensional space and $g(\vec{r}_i^a - \vec{r}_j^a)$ is the Gaussian normalized probability distribution s.t. $\int d\vec{r} \, g(\vec{r}) = 1$. After linearization over the $\sum_i \sigma_i \delta\left(\vec{x} - \vec{r}_i^\alpha\right)$ and putting into expression for $\langle Z^n \rangle_P$ value of distribution function $P\{\sigma\}$, we will be led to:

$$\langle Z^n \rangle_P \propto \int D\vec{r}_i^\alpha \, g(\vec{r}_i^\alpha - \vec{r}_{i+1}^\alpha) \exp(-C \sum_\alpha \int d\vec{x} \, \hat{\rho}_\alpha^3(\vec{x}) - B_0 \bigg/ \sum_\alpha \int d\vec{x} \, \hat{\rho}_\alpha^2(\vec{x})) \times$$
$$\int D\psi_\alpha(\vec{x}) \exp[1/2 \sum_{\alpha < \beta} \int d\vec{x} d\vec{y} \psi_\alpha(\vec{x}) \psi_\beta(\vec{y}) \hat{q}_{\alpha\beta}(\vec{x}, \vec{y}) - \qquad (2.7)$$
$$- \frac{1}{2|B|} \sum_\alpha \int d\vec{x} d\vec{y} \psi_\alpha(\vec{x}) \psi_\beta(\vec{y}) \delta(\vec{x} - \vec{y})]$$

where

$$\hat{\rho}_\alpha(\vec{x}) = \sum_i \delta(\vec{x} - \vec{r}_i^\alpha)$$
$$\hat{q}_{\alpha\beta}(\vec{x}, \vec{y}) = \sum_{ij} K_{ij} \delta(\vec{x} - \vec{r}_i^\alpha) \delta(\vec{x} - \vec{r}_j^\beta) \qquad (2.8)$$

Thus

$$\langle Z^n \rangle_P \propto \int D\rho_\alpha Dq_{\alpha\beta} \exp[-F(\rho_\alpha, q_{\alpha\beta})] \qquad (2.9)$$

where $F(\rho_\alpha, q_{\alpha\beta}) = E(\rho_\alpha, q_{\alpha\beta}) - S(\rho_\alpha, q_{\alpha\beta})$ is the free energy functional, $E(\rho_\alpha, q_{\alpha\beta})$ is conformational energy, $S(\rho_\alpha, q_{\alpha\beta})$ - entropy of $n$ polymer chains which corresponded to polymer chains residues densities $\{\rho_\alpha\}$ and two-replica overlap parameters $\{q_{\alpha\beta}\}$ :

$$-E(\rho_\alpha, q_{\alpha\beta}) = -C\sum_\alpha \int_{\vec{x}} d\vec{x}(\hat{\rho}_\alpha(\vec{x}))^3 - B_0 \sum_\alpha \int dx \left(\hat{\rho}_\alpha^2(x)\right) + \ln \int D\psi_\alpha(\vec{x}) \times$$

$$\times \exp\left\{\frac{1}{2}\sum_{\alpha\beta}\iint_{xy} d\vec{x}d\vec{y}\psi_\alpha(\vec{x})\psi_\beta(\vec{y})q_{\alpha\beta}(\vec{x},\vec{y}) - \frac{1}{2|B|}\sum_\alpha \iint_{xy} d\vec{x}d\vec{y}\psi_\alpha(\vec{x})\psi_\beta(\vec{y})\delta(\vec{x}-\vec{y})\right\}$$

(2.10)

$$\exp[S(\rho_\alpha, q_{\alpha\beta})] = \int D\vec{r}_i^\alpha g(\vec{r}_{i+1}^\alpha - \vec{r}_i^\alpha)\delta(\rho_\alpha - \hat{\rho}_\alpha)\delta(q_{\alpha\beta} - \hat{q}_{\alpha\beta})$$

In the mean-field approximation we need to minimize the free energy functional $F(\rho_\alpha, q_{\alpha\beta})$ over the one- and two-replica order parameters $\rho_\alpha, q_{\alpha\beta}$.

The expressions, which are obtained above are identical to them, obtained in [4] for the random sequence without correlations. The main difference is in two-replica overlap parameter $q_{\alpha\beta}$ definition (see eqn(2.7)).

Let us make Fourie-transformation of the order parameter of the system :

$$q_{\alpha\beta}(\vec{r}) = V^{-1}\sum_{\vec{k}\neq 0}q_{\alpha\beta}(\vec{k})\exp(i\vec{k}\vec{r}) \qquad (2.11)$$

where $V$ indicates volume used by macromolecule. This transformation will led us to a new expression for conformational energy :

$$-E(q) = \ln \int D\psi_\alpha(\vec{k})\exp\left\{-\frac{V}{2}\sum_{\alpha\beta}\sum_{\vec{k}\neq 0}\psi_\alpha(\vec{k})\psi_\beta(-\vec{k})\left[\frac{\delta_{\alpha\beta}}{|B|} - q_{\alpha\beta}(\vec{k})\right]\right\} (2.12)$$

Using Gaussian properties of this integral expression for conformational energy can be rewrite as :

$$-E(q) = const - 1/2\sum_{\vec{k}\neq 0}\ln\det P_{\alpha\beta}(\vec{k})$$

$$P_{\alpha\beta}(\vec{k}) = \frac{\delta_{\alpha\beta}}{|B|} - q_{\alpha\beta}(\vec{k}) \qquad (2.13)$$

Microscopic order parameter of the system can be displayed in the following form :

$$\hat{q}_{\alpha\beta}(\vec{x}-\vec{y}) = \hat{Q}_{\alpha\beta}(\vec{x}-\vec{y}) + \sum_{i\neq j}K_{ij}\delta(\vec{x}-\vec{r}_i^\alpha)\delta(\vec{y}-\vec{r}_j^\beta) \quad (2.14)$$

where $\hat{Q}_{\alpha\beta}(\vec{x}-\vec{y})$ is two-replica overlap parameter, which was used in some articles dedicated to random heteropolymers with non-correlated sequences [3,4]:

$$\hat{Q}_{\alpha\beta}(\vec{x}-\vec{y}) = \sum_i \delta(\vec{x}-\vec{r}_i^{\alpha})\delta(\vec{y}-\vec{r}_i^{\beta}) \quad (2.15)$$

From the normalization condition

$$\int \hat{Q}_{\alpha\beta}(\vec{x}-\vec{y})d\vec{x} = \rho_{\alpha} \quad (2.16)$$

the order parameter $\hat{Q}_{\alpha\beta}(\vec{x}-\vec{y})$ had been found as it was shown by Shakhnovich and Gutin [3]

$$\hat{Q}_{\alpha\beta}(\vec{x}-\vec{y}) = \frac{\rho}{R^d}\varphi_{\alpha\beta}(\frac{\vec{x}-\vec{y}}{R}) \quad (2.17)$$

where R is the characteristic scale of two-replica overlap and

$$\int d\vec{z}\,\varphi_{\alpha\beta}(\vec{z}) = 1 \quad (2.18)$$

Order parameter of the system $q_{\alpha\beta}$, obviously satisfying to normalization conditions as followed :

$$\int d\vec{y}\, q_{\alpha\beta}(\vec{x},\vec{y}) = \overline{K}\rho_{\alpha}(\vec{x})$$
$$\overline{K} \equiv \sum_j K_{ij} \quad (2.19)$$

Because the thermal average of quantity $\delta(\vec{x}-\vec{r}_i^{\alpha})\delta(\vec{y}-\vec{r}_j^{\beta})$ may be interpret as *aprior* probability of corresponding localization of $\alpha$ replica i-th residue and $\beta$ replica j-th residue ( $\Pr ob = P_{ij}^{\alpha\beta}(x,y)$ ), the order parameter $q_{\alpha\beta}(k)$ necessary to find in the following form

$$q_{\alpha\beta}(\vec{k}) = (K\rho/R^d)\varphi_{\alpha\beta}(\frac{\vec{x}-\vec{y}}{R}) + \sum_{i \neq j} K_{ij} P_{ij}^{\alpha\beta}(\vec{x},\vec{y}) \quad (2.20)$$

Taking into account the system translation invariance probability distribution $P_{ij}^{\alpha\beta}(x,y)$ necessary to find as

$$P_{ij}^{\alpha\beta}(x,y) = Const P_{ij}^{\alpha\beta}(y|x) \quad (2.21)$$

where $P_{ij}^{\alpha\beta}(y|x) = \Pr ob \{$ $\beta$ replica j-th residue situated in point $y$, if the $\alpha$ replica i-th residue situated in point $x$ $\}$ is conditional probability distribution.

It is known [15] that polymer chain behavior in globular state described by Gaussian statistic .

According to this by analogy with [3,4] and taking into account the normalization condition (eqn. 2.19) we will use for $q_{\alpha\beta}$ the following equation :

$$q_{\alpha\beta}(\vec{x}-\vec{y}) = KQ_{\alpha\beta}(\vec{x}-\vec{y}) + \sum_{i \neq j} K_{ij} P_{ij}^{\alpha\beta}(\vec{x}-\vec{y}) \tag{2.22}$$

$$P_{ij}^{\alpha\beta}(\vec{x}-\vec{y}) = \int dz P(\vec{z}-\vec{x}, |j-i|) \frac{\rho}{NR^d} \varphi_{\alpha\beta}(\frac{\vec{z}-\vec{y}}{R})$$

where

$$P(\vec{x}-\vec{z}, |j-i|) \propto (a^2|j-i|)^{-d/2} \exp\left(-\frac{(\vec{x}-\vec{z})^2}{a^2|j-i|}\right) \tag{2.23}$$

Here $P(r, |j-i|)$ is the probability distribution of the end-to-end vector $r$ for the Gaussian polymer chain. After simplifying in the limit $N \to \infty$ we will come to a newer one equation for Fourie-transformation of the order parameter:

$$\begin{aligned}\tilde{q}_{\alpha\beta}(\vec{k}) &= \frac{\rho}{a^d} \tilde{\varphi}_{\alpha\beta}(\vec{k}R) \frac{1}{N} \sum_i \sum_{j \neq i} K(|j-i|) \exp\{-a^2 k^2 |j-i|\} \cong \\ &\cong K\rho\tilde{\varphi}_{\alpha\beta}(\vec{k}R) + 2\rho \sum_{l \geq 1} K(l) \exp(-a^2 k^2 l) \tilde{\varphi}_{\alpha\beta}(\vec{k}R) \equiv \rho\tilde{\varphi}_{\alpha\beta}(\vec{k}R) A(\vec{k})\end{aligned} \tag{2.24}$$

where $k = |\vec{k}|$. It is obvious that without correlations in polymer chain sequence our results reduce to obtained recently in [4].

Using results of [16,4] it will lead us to the following form of conformational energy, in limit $n \to 0$ :

$$\frac{E}{n} \cong -1/2 \sum_k \int_0^1 \frac{dx}{x^2} \ln \lambda_k(x) \tag{2.25}$$

where

$$\lambda_k(x) = 1/|B| - \rho A(k) - \int_x^1 dy M_k(y) - x M_k(x) \tag{2.26}$$

and the Parisi function $M_k(x)$ parametrising the off-diagonal elements of the hierarchical matrix $P_{\alpha\beta}(k)$ in the $n \to 0$ limit.

Now we have to minimize the free energy functional (see eqn. (2.8, 2.9)). It is known [3,4], that replica-symmetric solution is invalid for random heteropolymers, because in this case the entropy contribution has form[3]:

$$S\{\rho_\alpha, q_{\alpha\beta}\} = (n-1)\Delta S(R) \quad (2.27)$$

where $\Delta S(R)$ is the loss of entropy of ideal polymer chain constrained in tube of diameter $R$. Here

$$\Delta S(R) \cong \begin{cases} -Na^2/R^2 & (R \gg a) \\ Nd\ln(R/a) & (R \ll a) \end{cases} \quad (2.28)$$

Following the Parisi ansatz for one-step symmetry breaking, for $n$ replicas, there are $n/x$ groups with $x$ replicas per group. The entropy loss is therefore:

$$S\{\rho_\alpha, q_{\alpha\beta}\} = \frac{n}{x}(x-1)\Delta S(R) \quad (2.29)$$

In the same one step R.S.B. the energy contribution is come to the following form (for $n \to 0$)

$$E = \frac{E\{q_{\alpha\beta}\}}{n} = T/2 \int \frac{d\vec{k}}{R^d} \left\{ \frac{1}{x} \ln\left[b - x\rho A(\vec{k}/R)(1-\tilde{\varphi}(\vec{k}))\right] + \left(1 - 1/x\right)\ln b \right\} \quad (2.30)$$

where $b = 1/|B|$

For the subsequent minimization consider the above defined (see eqn.(2.23)) function $A(\vec{k})$ behavior

$$A(\vec{k}/R) = K + 2\sum_{l \geq 1} K(l)\exp(-l\frac{a^2 k^2}{R^2}) \quad (2.31)$$

For the correlation function (2.4) the $A(\vec{k}/R)$ can be evaluated as

$$A(\vec{k}/R) \cong K\left\{1 + 2\Gamma(\beta)\left(\frac{ak}{R}\right)^{-2\beta}\right\} \quad \text{for } ak/R \ll 1 \quad (2.33)$$

Consequently, the function $A(\vec{k}/R)$ is increased monotonically with $R/k$ and thus for the any $k$ value we can find the $R$ scale s.t. the energy contribution in eqn(2.30) will be diverge as $\ln(b - const(R/a)^{2\beta})$. From the form of eqns(2.30) we can see that there is a wide scale $R$ region, which is forbidden energetically, because, the dependence $E$ vs. $R$ have form presented in fig.1a. The entropy loss $\Delta S(R)$ due by restrictions with scale $R$ has form presented in fig.1b. It is obvious that for the small enough values of **R** scale we have the situation identical to the [3,4] ( for the RHP with non-correlated sequence ). Consequently , free energy for the one-step R.S.B. have one maximal value at the $R \approx 0$ (see fig. 1c) and , correspondingly , one thermodynamically stable state. Moreover, the free energy for the one-step R.S.B. have the infinite energetic barrier due by above mentioned divergence, which separated the stable state with $R \approx 0$ from the other (unfrozen) states.

In the case of two-step R.S.B. we carried out the calculations analogous to eqn (2.30). For this scheme of replica symmetry braking the conformational energy is diverge at enough large replicas overlap scale too. Thus, the described above results does not due by one-step R.S.B. approximation and the examined here system properties are reflected.

### 3.Discussion.

Alongside with some peculiarities that were investigate there appear in the heteropolymer chain folding in presence of long-range correlations in residues' sequence. The inter-residual correlations defines by Gaussian distribution function (2.2) with non-diagonal correlation matrix (2.3). In difference of non-correlated sequence case this probability distribution doe not factored on the terms corresponded to the chain residues. In the present paper the standard techniques, developed in [3] were use. The results that were obtained above are mathematically similar to them in [4] , but two-replica order parameter redefinition led to unexpected physical properties of examined system.

In the case of power-law correlations decay (see eqn.(2.4)) the off-diagonal terms which have contribution in the order parameter $q_{\alpha\beta}$ (see eqn.(2.8)) led to the qualitative different behavior of energetical term (2.30). Taking into account estimation (2.33) we can see that each harmonic $\vec{k}$ energetic contribution characterized by some space scale $R$ of divergence, defined by following expression

$$b - x\rho A(\vec{k}/R)\left(1 - \widetilde{\varphi}(\vec{k})\right) = 0 \qquad (3.1)$$

At the large enough value of $R$ the last expression may be rewrite as

$$b \cong 2x\rho\left(1 - \tilde{\varphi}(\vec{k})\right)\Gamma(\beta)\left(\frac{R}{ak}\right)^{2\beta} \quad (3.2)$$

Expressions (2.17,18) show that the characteristic scale of the $\varphi(\vec{z})$ function decay is $|\vec{z}| \approx 1$ and $\varphi(\vec{z}) = \varphi(|\vec{z}|)$ has form, schematically represented in fig.2. Consequently, exists $k^*$, s.t. for any $k > k^*$ $\tilde{\varphi}(\vec{k}) < 1$. So, for example, for $\varphi(r) \propto \exp(-r^2/2)$ we have $\tilde{\varphi}(\vec{k}) \propto \exp(-k^2/2)$.

Thus, for $k > k^*$ eqn.(3.2) have solution, defined by expression

$$R \cong ak\left(\frac{b}{2\rho x\left(1 - \tilde{\varphi}(\vec{k})\right)}\right)^{1/(2\beta)} \quad (3.3)$$

and the system energy may be considered as superposition of different harmonics contributions, as represented in fig.3. It is obvious that for the large enough values of $R$ scale the system energy became diverge. Thus, for the polymer globule the frozen state (with $R \approx 0$) is stable only. Taking into account the normalization condition for $g(\vec{r})$ (see explanations to eqn. (2.6) ), the random coil state free energy in the polymer globule mean-field theory [15] evaluated as $F = 0$. Consequently, our system has two stable states. One, corresponded to the frozen chain backbone path and another, random coil state. Because in the large enough values of $R$ scale the system energy became diverge these stable states are separate by infinite energetic barrier. This result may be interpreted as follows. The heteropolymer chain with long-correlated sequence can exist in folded $R \approx 0$ or random coil state, but the folded state is kinetically do not available. Thus, the power-law correlations led to the random heteropolymer folding impossibility.

It's interesting that the correlations exponential decay

$$K(l) \propto \exp(-l/\xi) \quad (3.4)$$

in the limit $\xi \gg 1$ is equivalent to $\beta = 1$ in (2.4) which is the marginal case of maximal long-range correlations in sequence

$$\tilde{q}_{\alpha\beta}(\vec{k}) = \rho\tilde{\varphi}_{\alpha\beta}(\vec{k}R)\left\{1 + 2\sum_{l\geq 1}\exp\left[-l\left(a^2k^2 + 1/\xi\right)\right]\right\} \cong$$
$$\cong \rho\tilde{\varphi}_{\alpha\beta}(\vec{k}R)\left\{1 + 2\sum_{l\geq 1}\exp\left(-a^2k^2 l\right)\right\} \quad (3.5)$$

If we have $\xi \approx 1$, then

$$\tilde{q}_{\alpha\beta}(\vec{k}) \cong \rho \tilde{\varphi}_{\alpha\beta}(\vec{k}R)\left\{1 + 2\sum_{l\geq 1}\exp(-l/\xi)\right\} \tag{3.6}$$

which is completely equivalent to the order parameter obtained in [4] for the heteropolymer with non-correlated sequence. It's quite natural because in the case of $\xi \approx 1$ the chain sequence may be divide into enough small pieces that will be statistically independent.

The above obtained results are in agreement with hypothesis about long-range correlations in non-coding DNA sequences only [8]. Recently, Shakhnovich and Gutin[7] found that for RHP to have the energetically stable folded state it is sufficient that monomers' sequence forms an uncorrelated random sequence. Our results show that the lack of sequence long range correlations is a necessary condition for RHP folding possibility.

**References:**

1.  B.Derrida (1980) Phys. Rev. Lett. 49 ,79

2.  J.B. Bryngelson & P.G.Wolynes (1987) Proc. Natl. Sci. USA 84 ,7524

3.  E. Shakhnovich & A. Gutin (1989) Biophys.Chem. 34 , 187

4.  C. Sfatos , A. Gutin , E. Shakhnovich (1993) Phys.Rev. E 48 ,465

5.  T. Garel & H. Orland (1988) Europhys. Lett. 6 ,597

6.  C. Sfatos , A. Gutin , E. Shakhnovich (1994) Phys.Rev. E 50 , 2898

7.  E. Shakhnovich & A. Gutin (1990) Nature 346 , 773

8.  C.-K. Peng, S.V.Buldyrev, A.L.Goldberger, S.Havlin, F.Sciortino, M.Simons & H.E.Stanley (1992) Nature 356 , 168

9.  W.Li & K.Kanelko (1992) Europhys. Lett. 17 ,655

10. R.Woss (1992) Phys.Rev. Lett. 68, 3805

11. S.Nee (1992) Nature 357 , 450

12. V.V.Prabhu & J.-M. Claverie (1992) Nature 357 , 782

13. A. Gutin , C. Sfatos & E. Shakhnovich (1994) J.Phys.A 27 , 7957

14. H.Stanley, S.V.Buldyrev, A.L.Goldberger, Z.G.Goldberger, S.Havlin, R.N.Mantegna, S.M.Ossadnik, C.-K.Peng & M.Simons (1994) Physica A 205 , 214

15. A.Yu. Grossberg & A.R.Khokhlov Statistical Physics of Macromolecules (AIP, New York ,1994)

16. M.Mezard & G.Parisi (1991) J. Physique I 1 , 809

**Figures captures**

**Fig.1a.** Conformational energy dependence vs. scale of replicas overlap ($R$). Here $v$ is the excluded volume of chain residue and $R_0$ is the scale of conformational energy divergence.

**Fig.1b.** Conformational entropy plotted vs. scale of replicas overlap ($R$). Here $v$ is the excluded volume of chain residue and $R_0$ is the scale of conformational energy divergence.

**Fig.1c.** Solid line is the free energy dependence vs. scale of replicas overlap ($R$). Dashed line is the free energy plotted vs. scale $R$ for the RHP with non-correlated sequence of residues (see, e.g. [3,4] ).

**Fig.2.** Overlap function behavior plotted vs. dimensionless scale of replicas overlap.

**Fig.3.** The scheme of different harmonics contributions for the conformational energy. $R_k$ is the scale of divergence for the wave vector length $k$ (see eqn(3.3)).

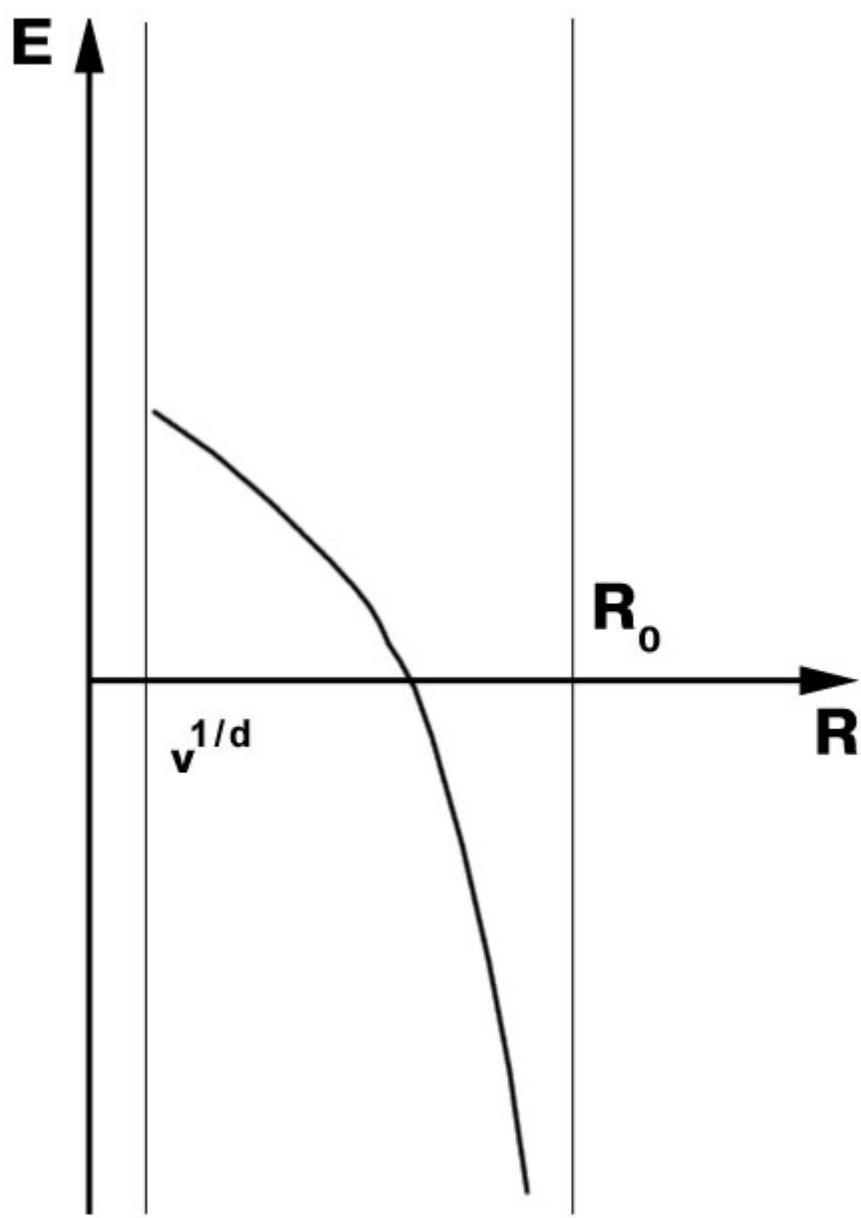

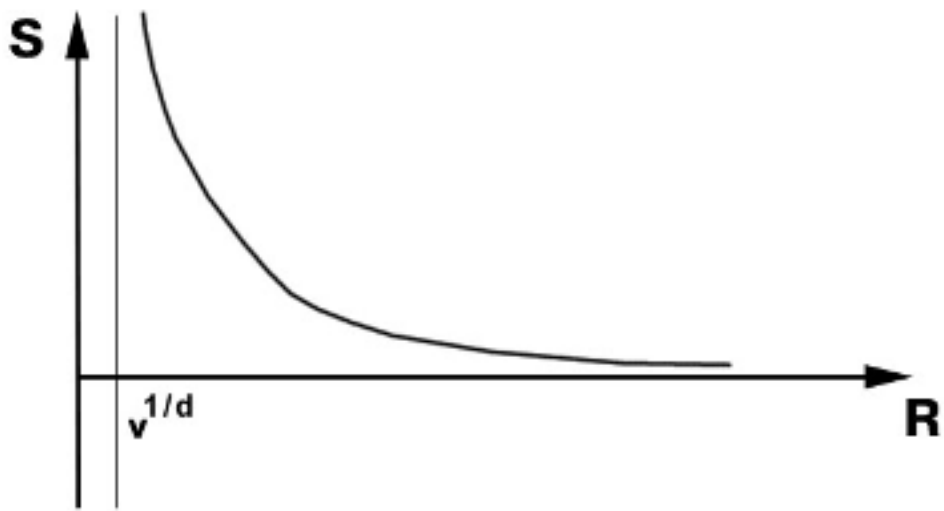

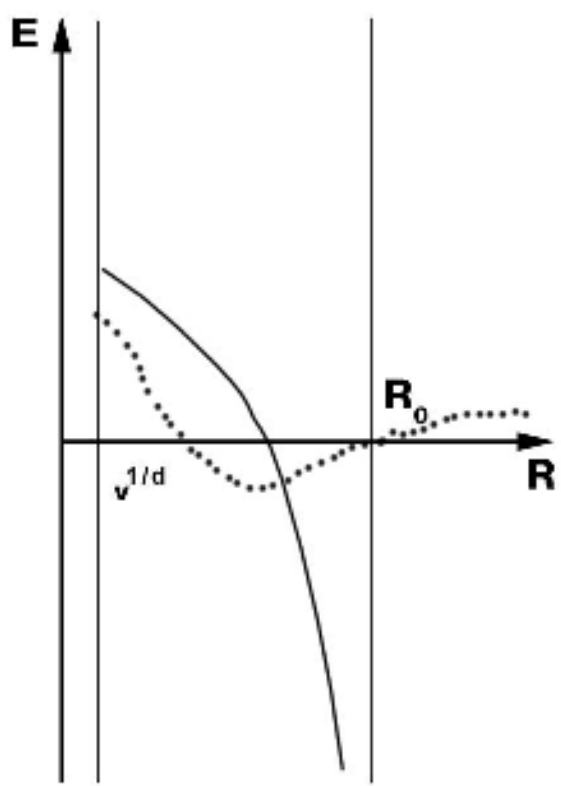

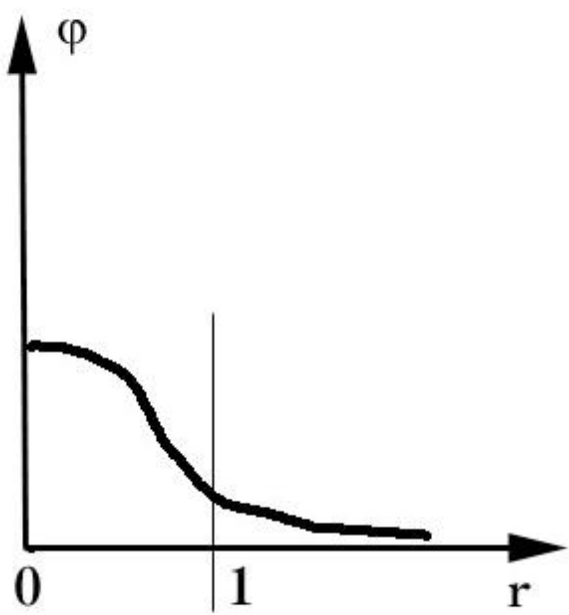

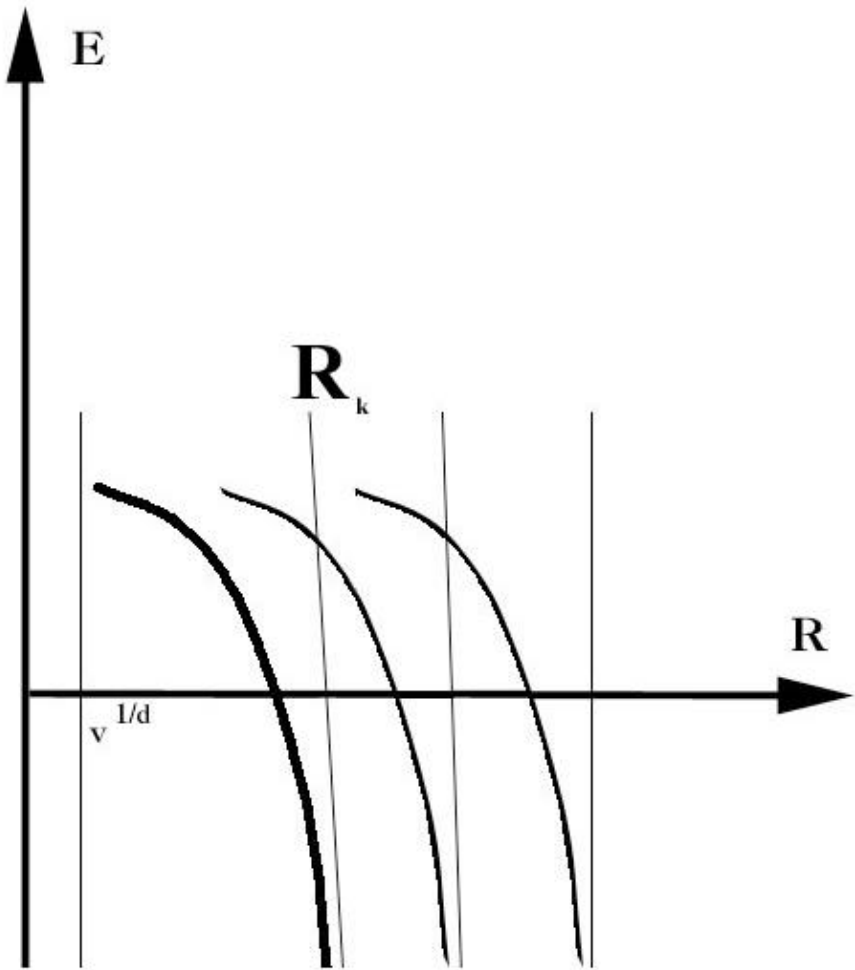